\documentclass[11pt,twoside]{article}
\usepackage{asp2010}

\resetcounters

\begin{document}

\title{IC 5181: An S0 Galaxy with Ionized Gas on Polar Orbits}

\author{A. Pizzella,$^{1,2}$ L. Morelli,$^{1,2}$ E.~M. Corsini,$^{1,2}$
  E. Dalla Bont\`a,$^{1,2}$ and\\ M. Cesetti$^1$}
\affil{$^1$Dipartimento di Fisica e Astronomia `G. Galilei',
  Universit\`a di Padova, Padova, Italy} 
\affil{$^2$INAF-Osservatorio Astronomico di Padova, Padova, Italy}

\begin{abstract}
The nearby S0 galaxy IC~5181 is studied to address the origin of the
ionized gas component that orbits the galaxy on polar orbit.  We
perform detailed photometric and spectroscopic observations measuring
the surface brightness distribution of the stars ($I$ band), ionized
gas of IC~5181 (H$\alpha$ narrow band), the ionized-gas and stellar
kinematics along both the major and minor axis, and the corresponding
line strengths of the Lick indices.  We conclude that the galaxy hosts
a geometrically and kinematically decoupled component of ionized
gas. It is elongated along the galaxy minor axis and in orthogonal
rotation with respect to the galaxy disk.  The result is suggesting
that the gas component is not related to the stars having an external
origin. The gas was accreted by IC~5181 on polar orbits from the
surrounding environment.
\end{abstract}

\section{Introduction}

IC~5181 is a large \citep[$2\farcm6\, \times\, 0\farcm8$;][hereafter
  RC3]{1991rc3..book.....D} and bright ($B_T=12.51$; RC3) early-type
disk galaxy. It is classified as edge-on SA0 in RC3 due to the
presence of a flattened thick disk. Its total absolute magnitude is
$M^0_{B, {\rm T}} = -19.48$ corrected for inclination and extinction
(RC3) and adopting a distance of 24.8 Mpc
\citep{1988ngc..book.....T}. IC~5181 is a member of the loose NGC~7213
group \citep{1989ApJS...69..809M, 1993A&AS..100...47G}. It forms a
pair with the edge-on spiral galaxy NGC~7232A at $8.1'$ separation
corresponding to a projected linear distance of 58.4 kpc.

In the framework of acquisition events occurred in the lifetime of
galaxies, we present IC~5181 as a new case of a disk galaxy
characterized by a geometric and kinematic orthogonal decoupling
between its stellar body and the ionized-gas component.

\section{Observations}

\begin{figure}[t!]
\centering
\includegraphics[angle=0.0,width=0.46\textwidth]{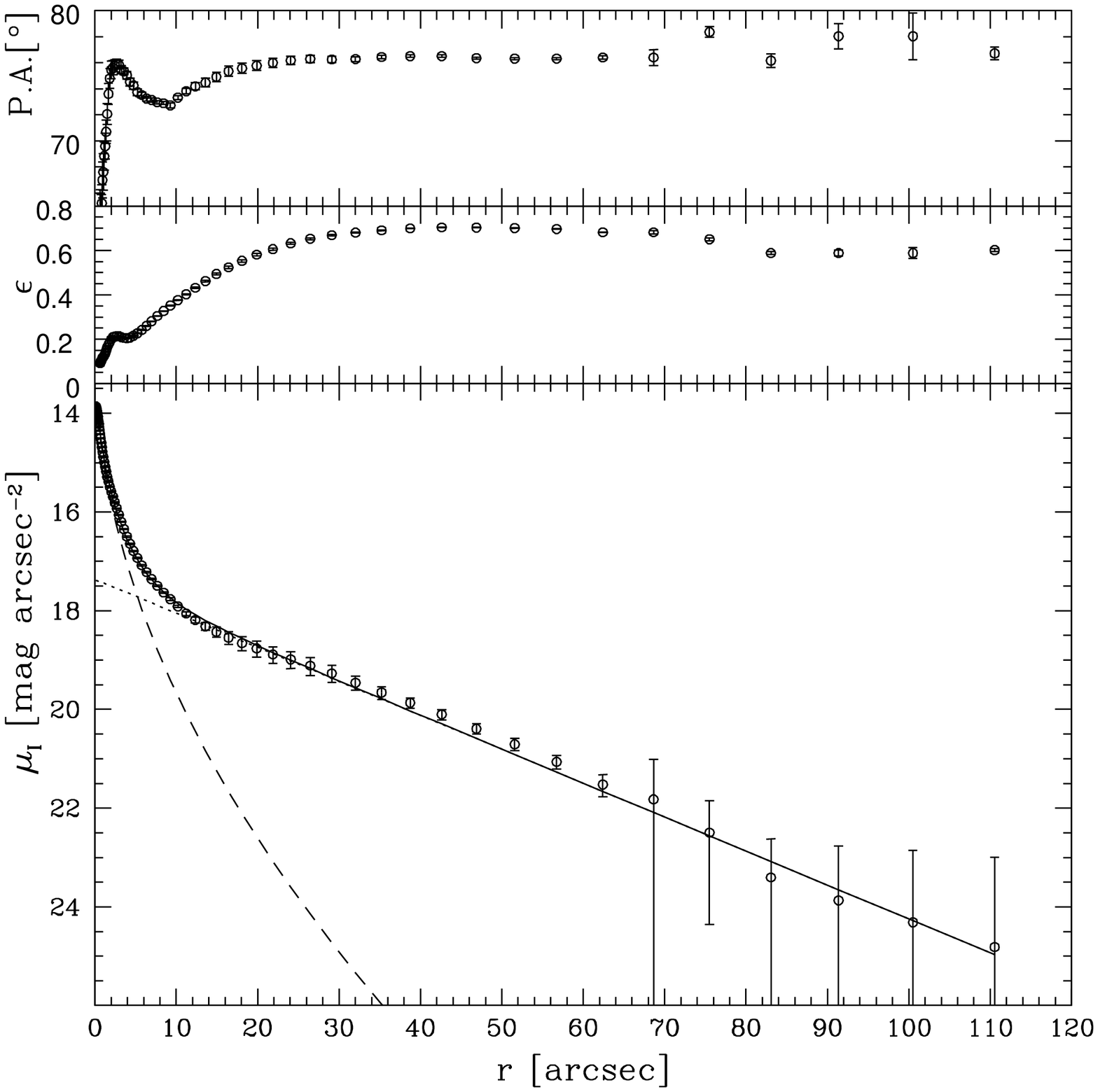}
\includegraphics[angle=0.0,width=0.46\textwidth]{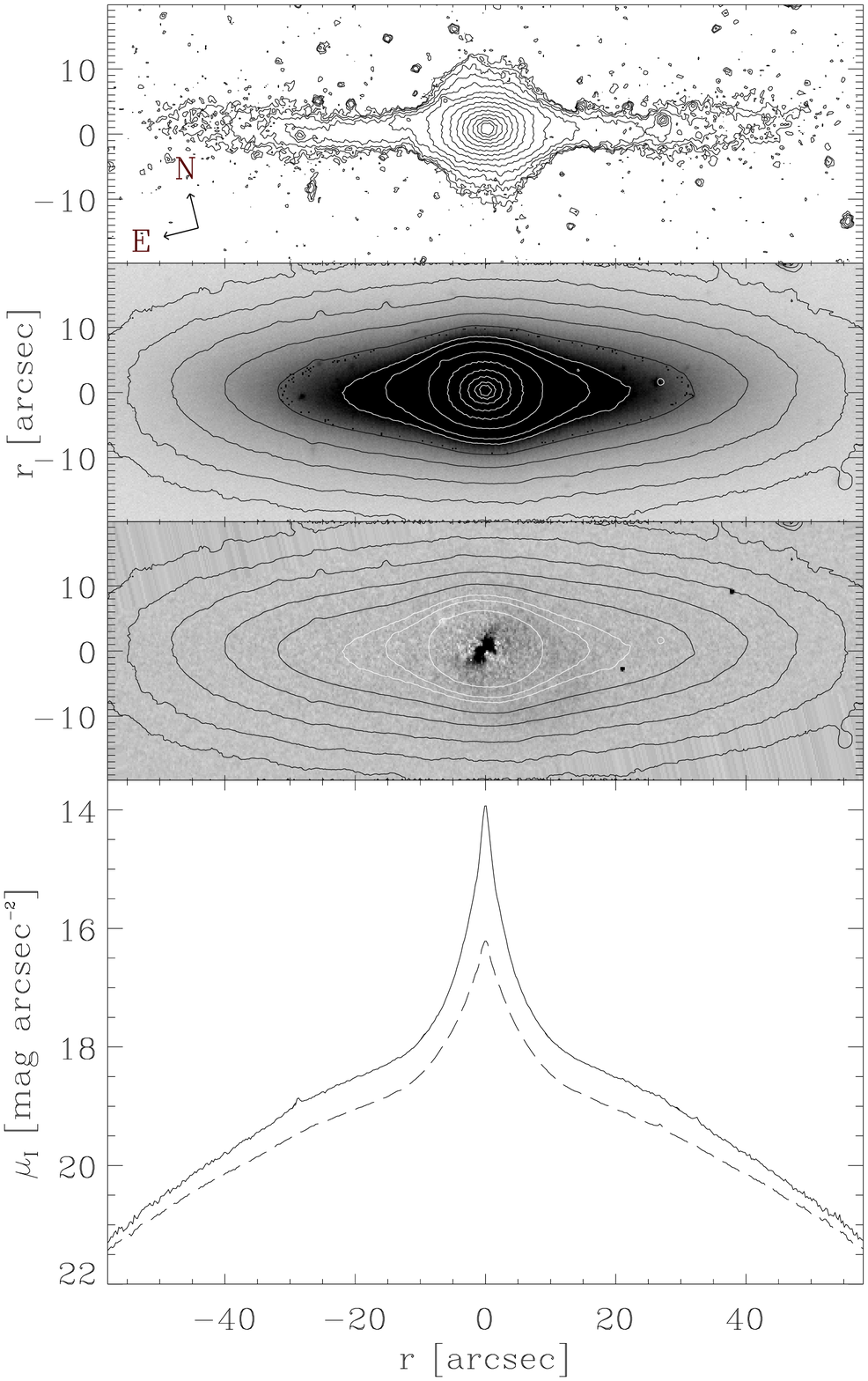}
 \caption{Left panel: The radial profiles of position angle,
   ellipticity, and Cousins $I$-band surface brightness as a function
   of the semi-major-axis distance are plotted (from top to
   bottom). The dashed, dotted, and solid lines represent the
   surface-brightness radial profiles of the bulge, disc, and model
   obtained from the photometric decomposition, respectively. Right
   panel: the unsharp-masked image, the continuum-band image with some
   $I-$band isophotal contours, the continuum-free
   H$\alpha$$+$[\ion{N}{II}] image with $I-$band isophotal contours,
   and the radial profiles of the surface brightness extracted along
   the major axis (solid line) and over a rectangular aperture with a
   width of $40\arcsec$ parallel to the major axis and centered on the
   galactic nucleus (dashed line) are shown (from top to bottom). The
   orientation of the field of view is given in the top panel.
\label{fig:phot}}
\end{figure}

\begin{figure*}[t!]
\centering
\includegraphics[angle=0.0,width=0.49\textwidth]{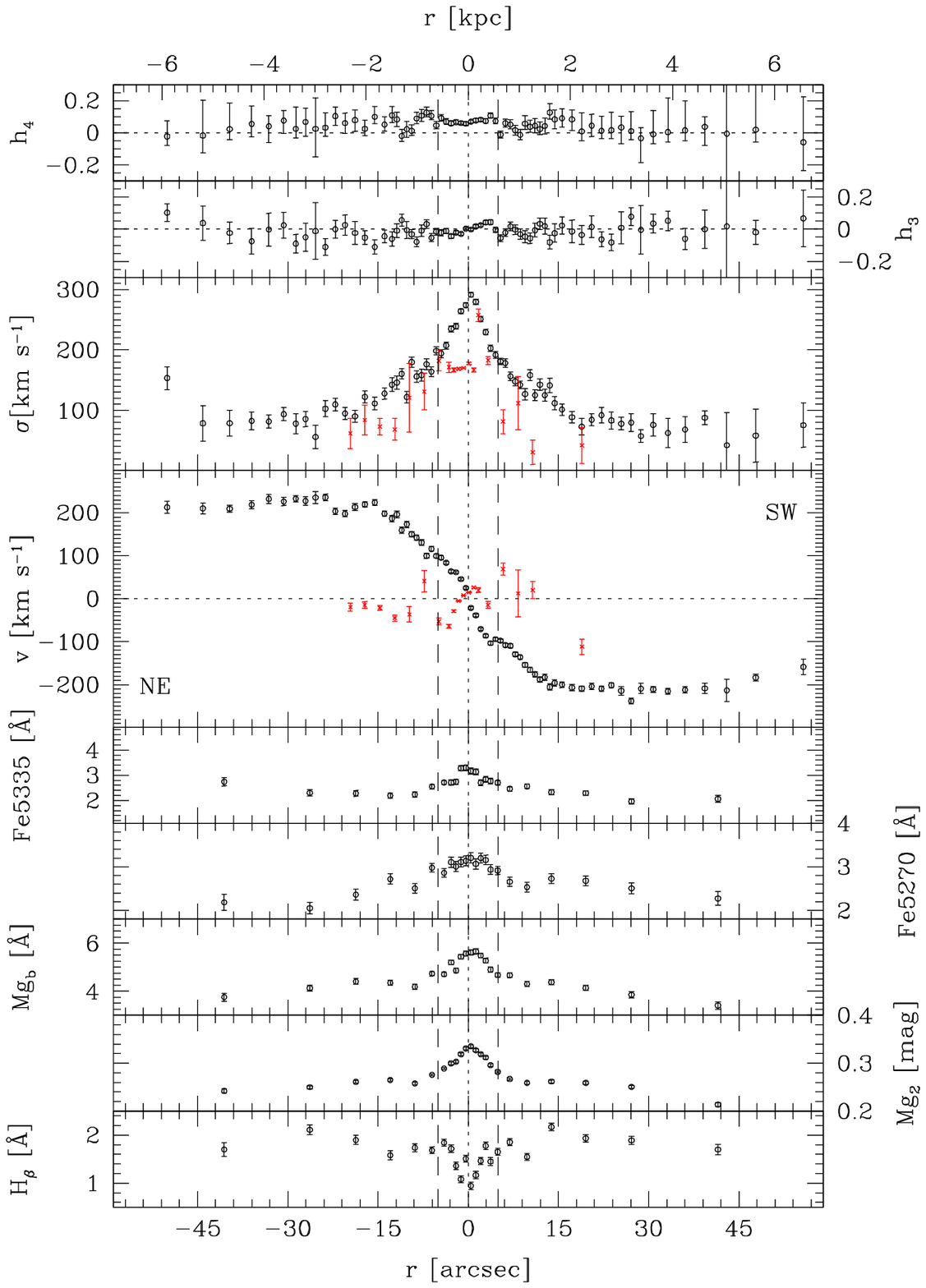}
\includegraphics[angle=0.0,width=0.49\textwidth]{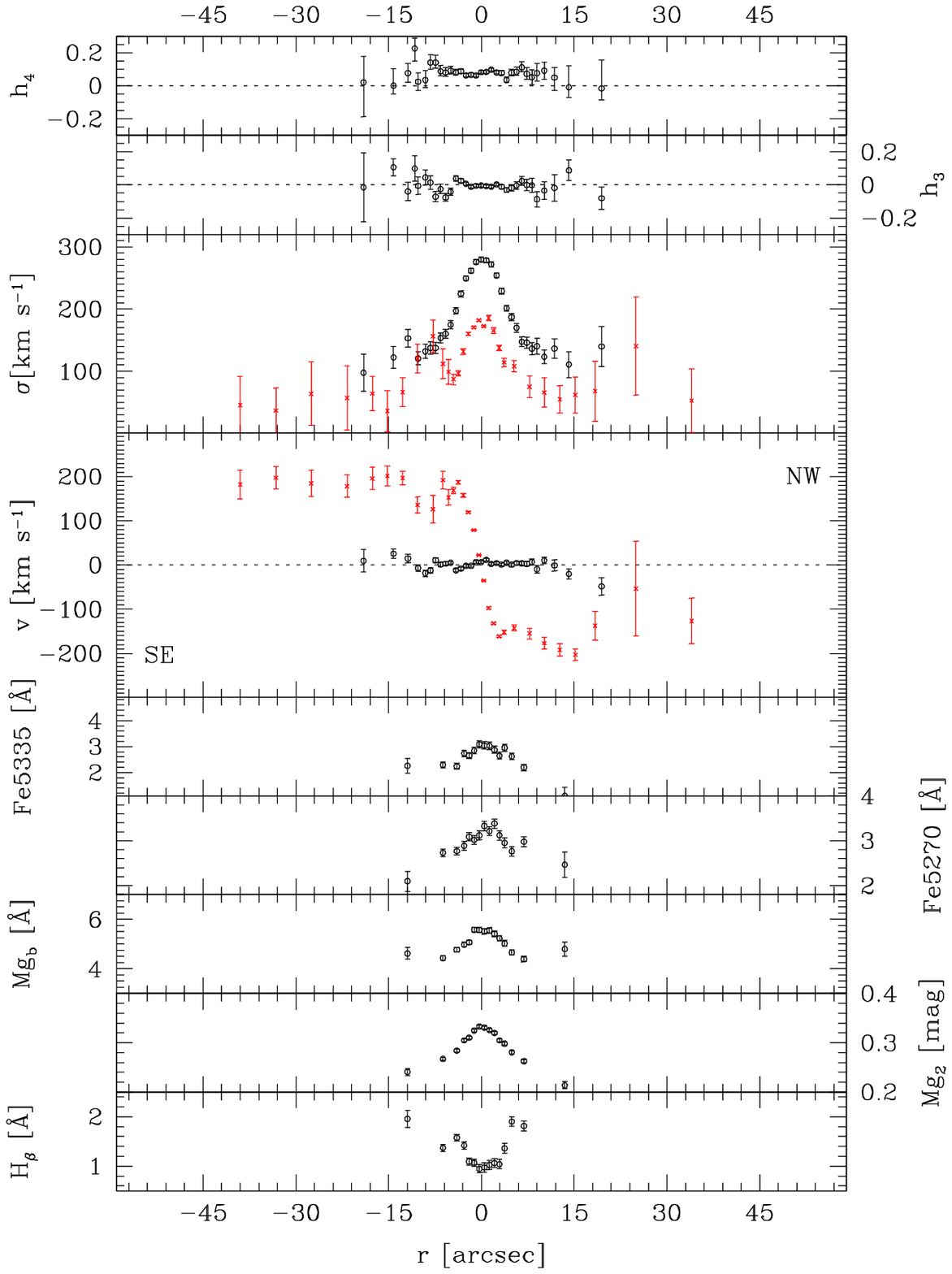}
 \caption{Left panel: Kinematic parameters of the stars (circles) and
   ionized gas (crosses) and the line-strength indices measured along
   the major axis of IC~5181 ($\rm PA = 74\deg$). The radial profiles
   of Gauss-Hermite cofficients $h_4$ and $h_3$, velocity (after the
   subtraction of systemic velocity), velocity dispersion, and
   line-strength indices Fe5335, Fe5270, Mg$b$ , Mg$_2$, and H$\beta$
   are plotted (from top to bottom). The vertical dashed lines
   correspond to the radii ($|R| = R_{\rm bd}$), where the
   surface-brightness contributions of the bulge and disk are
   equal. Right panel: As the left panel for the minor axis ($\rm PA =
   164\deg$).
\label{fig:kin}}
\end{figure*}

The photometric observations of IC~5181 were carried out at the
European Southern Observatory (ESO) in La Silla with the 2.2-m MPG/ESO
telescope equipped with the Wide Field Imager (WFI) on July 22, 1999.
The $I$-band and H$\alpha$-narrow band filters have been used.  The
spectroscopic observations were carried out at ESO in La Silla with
the 1.52-m ESO telescope on June 9-11, 1999 and the telescope was
equipped with the Cassegrain Boller \& Chivens spectrograph.  The
complete description of the observations can be found in
\citet{2013A&A...560A..14P}.  In Fig.~\ref{fig:phot} and
Fig.~\ref{fig:kin} we show the results of the photometric and
spectroscopic studies, respectively.

\section{Results}

The H$\alpha$ image of the galaxy shows an ionized gas structure
perpendicular to the stellar disk of IC~5181. The disk and bulge are
featureless and regular as expected for an S0 galaxy. The only
noticeable fact is the change in position angle ($\rm PA$) of the
isophotes around $5\arcsec< R <15\arcsec$ and that the disk is edge-on
($i_{\rm disk}=77\deg$).  The kinematics of the stellar component show
no anomalies. The ionized gas is instead moving on polar orbits and
its extension along the minor axis is twice the extension along the
major axis.

\begin{figure}[t!]
\centering
\includegraphics[angle=0.0,width=0.45\textwidth]{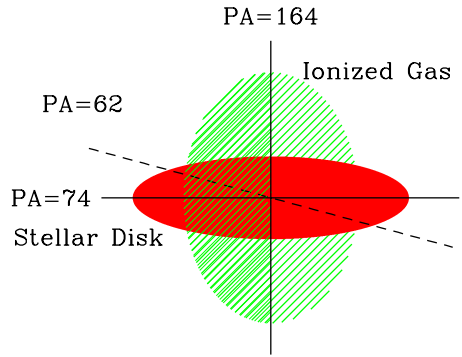}\\
\caption{Geometrical model of IC 5181. The stellar disk (red circle)
  has an inclination $i_{\rm disk}=77\deg$ and is oriented along $\rm
  PA=74\deg$.  The ionized gas polar disk (green circle) is
  perpendicular to the stellar disk and has an inclination of
  $41\deg$.  The ionized gas polar axis (dashed line) is aligned with
  the central triaxial structure along $\rm PA_{\rm gas} = 62\deg$,
  only $12\deg$ clockwise from the galaxy major axis.
\label{fig:modgeo}}
\end{figure}

We interpret these results as the indication that IC~5181 has a
central triaxial structure (a bar or a bulge) and that the ionized gas
is moving onto a plane perpendicular to its long axis.  To check if
this picture is consistent with the observations, we build a geometric
model of the galaxy. We derived its asymptotic circular velocity from
the stellar kinematics finding $v_{\rm circ} = (305\pm10)$ $\rm
km\;s^{-1}$.  We measure a projected asymptotic velocity of the
ionized gas of $v_{\rm gas} = (201\pm6)$ $\rm km\;s^{-1}$
(Fig.~\ref{fig:kin}, right panel). Assuming that the ionized gas is
moving with the same $v_{\rm circ}$ as the stars, we derive its
inclination: $i_{\rm gas} = 41\deg\pm3\deg$. The two angles describing
the orientation of the galaxy are therefore known: the inclination of
the stellar disk $i_{\rm disk}=77\deg$ and the inclination of the
ionized gas $i_{\rm gas} = 41\deg$. The direction of the angular
momentum of the gas can be found and it lies along $\rm PA_{\rm gas} =
62\deg\pm2\deg$. This is consistent with the orientation of the
ionized gas seen in the H$\alpha$ image (Fig.~\ref{fig:phot}, right
panel).  If the gas is orbiting in the equilibrium plane perpendicular
to the long axis of a triaxial structure, $\rm PA_{\rm gas}$ is also
the position angle of this structure
\citep[e.g.,][]{1985ApJ...292L..51B}. A scheme of the geometric model
can be found in Fig.~\ref{fig:modgeo}.  The small difference ($\Delta
{\rm PA}=12\deg$) between the position angles of the disk major axis
and bulge/bar long axis and the high inclination have prevented to
decompose it properly in a two-dimensional surface photometry
analysis.

The kinematical decoupling between the gaseous and stellar components
suggests the occurrence of an accretion event or merging
\citep{1999IAUS..186..149B}.  Therefore, it is straightforward to
explain the existence of the orthogonally rotating gas in IC~5181 as
the end result of the acquisition of external gas by the pre-existing
galaxy.  The nearby environment of IC~5181 shows no strong evidence of
such an event. IC~5181 does not interact with its closest companion,
NGC~7232A, and is one of the few galaxies of the NGC~7213 group to be
undetected in \ion{H}{I} \citep{2001MNRAS.324..859B}.
\citet{2001A&A...374...83B} proved that the environment of galaxies
that experienced past gas accretion do not appear statistically
different from those of normal galaxies. In addition, the polar
orientation of the ionized gas and its possible low metal content fits
well with the scenario proposed for the formation of polar ring
galaxies, where the gaseous ring is formed by the accretion of
material from a cosmic filament \citep{2006ApJ...636L..25M,
2008ApJ...689..678B, 2012MNRAS.425.1967}.

\acknowledgements The work was supported by Padua University through
grants 60A02-1283/10, 60A02-5052/11, and 60A02-4807/12. M. C. and
L. M.  thank financial support from Padua University grant
CPDR115539/11 and CPS0204, respectively.


\end{document}